\documentclass[twocolumn,pre,showkeys, showpacs]{revtex4}

\usepackage{graphicx} 
\renewcommand{\H}{$H$} \renewcommand{\S}{$S$}
\begin{document}

\title{Adaptation and enslavement in endosymbiont-host associations}

\author{Marcus R. Frean} \homepage{www.mcs.vuw.ac.nz/~marcus}
\affiliation{School of Mathematical and Computing Sciences, Victoria
  University, Wellington, New Zealand} \author{Edward R. Abraham}
\homepage{www.mcs.vuw.ac.nz/~abraham} \affiliation{National Institute
  of Water and Atmospheric Research (NIWA), P.O.  Box 14-901,
  Kilbirnie, Wellington, New Zealand }

\date{December 10, 2003; revised February 24, 2004}

\begin{abstract}
  The evolutionary persistence of symbiotic associations is a puzzle.
  Adaptation should eliminate cooperative traits if it is possible to
  enjoy the advantages of cooperation without reciprocating - a facet
  of cooperation known in game theory as the Prisoner's Dilemma.
  Despite this barrier, symbioses are widespread, and may have been
  necessary for the evolution of complex life.  The discovery of
  strategies such as tit-for-tat has been presented as a general
  solution to the problem of cooperation. However, this only holds for
  within-species cooperation, where a single strategy will come to
  dominate the population. In a symbiotic association each species may
  have a different strategy, and the theoretical analysis of the
  single species problem is no guide to the outcome.  We present basic
  analysis of two-species cooperation and show that a species with a
  fast adaptation rate is enslaved by a slowly evolving one.
  Paradoxically, the rapidly evolving species becomes highly
  cooperative, whereas the slowly evolving one gives little in return.
  This helps understand the occurrence of endosymbioses where the host
  benefits, but the symbionts appear to gain little from the
  association.
\end{abstract}

\keywords{endosymbiosis, mutualism, prisoner's dilemma, tit-for-tat,
  red king, miser}
\pacs{87.23.Kg}

\maketitle

\section{\label{sec:introduction} Introduction}

Cooperation between endosymbionts and their hosts is most striking in
the common case where the symbiont can survive in a free-living form
and each host is infected anew in its infancy \cite{Maynardsmith95,
  Frank96, Herre98}. If the host is unable to respond to individual
symbionts who do not contribute to the association, then free-riders
will prosper and the symbiont will ultimately become parasitic on its
host \cite{Bremermann86, Knolle89}. On the other hand, if the host is
able to punish defecting symbionts, the expectation is that a ``you
scratch my back, and I'll scratch yours'' relationship will develop
\cite{Trivers71, Axelrod84}. It is supposed that the reciprocation of
both cooperation and defection maintains a mutualistic relationship,
from which both parties benefit.  This appears to be supported by
game-theoretic analyses of simple models of cooperation, with
tit-for-tat strategies being successful in Prisoner's Dilemma
scenarios \cite{Axelrod81, Nowak92, Dugatkin98}. Although tit-for-tat
emerges as the winning strategy under a wide range of conditions, most
studies of cooperation are restricted to intra-specific competition.
It is not clear that this can be generalized to the host-endosymbiont
system, where the Prisoner's Dilemma occurs between two species.

In natural systems, it is found that many horizontally transmitted
non-obligate endosymbionts apparently generate large benefits for
their hosts. Examples include nitrogen fixation by rhizobial bacteria
in legumes \cite{Sprent93}; enhanced uptake of nutrients due to
mycorrhizal fungi \cite{Smith97}; carbon uptake by green algae in many
aquatic invertebrates, such as the zooxanthellae of corals
\cite{Veron95}; and bioluminescence provided by bacteria in squid and
fish \cite{Mcfallngai99}. Surprisingly, in very few cases have
endosymbionts been shown to benefit significantly from their
interactions with host organisms \cite{Douglas89}. For example,
\emph{Rhizobia} reproduce happily enough when free-living in the soil,
but most of these bacteria hardly reproduce at all once inside the
root nodules of legumes \cite{Sprent93, Denison00}.  For the putative
benefits of symbiotic life as zooxanthellae, dinoflagellates give up
their cell wall and their flagella, sacrifice most of their
photosynthetic products, and reduce their reproductive rate
\cite{Saffo01}. The non-obligate forms of endomycorrhizal fungi can be
readily grown alone, and there is no evidence for a fitness-related
benefit from symbiosis for them \cite{Smith97}. In particular, there
is no evidence that carbon flows from the plant to the fungus in
return for the nutrients these endosymbionts provide. Thus notions of
host and parasite appear to be turned on their head, and this
asymmetry is not explained by the standard tit-for-tat solution to the
problem of cooperation.

Most previous models of cooperation have only represented one
population.  In this paper we present a simple model of a two-species
association based on reciprocal altruism, and explore whether it can
explain both the maintenance of symbiotic relationships and the
apparent enslavement of the symbionts by their hosts. In keeping with
much previous work on the evolution of cooperation, we formalize the
Prisoner's Dilemma in terms of the costs, $c$, and benefits, $b$, of
cooperative acts. The species have variable degrees of cooperation,
depending on the value of these parameters \cite{Frean96, Wahl99a,
  Wahl99b, Killingback02}. The degree of cooperation adopted by each
species is subject to evolutionary change. The challenge is then to
demonstrate that a mutualistic relationship is stable in the face of
mutations in the parameters governing cooperation.

A core asymmetry between the host and the endosymbionts is that the
symbionts have a shorter generation time. This rapid turnover allows
their population to respond quickly to changes in the hosts' behavior.
A consequence of this differential adaptive rate is that the
tit-for-tat strategy is no longer optimal. Rather the endosymbiont
becomes fully cooperative and the host becomes miserly, adopting a
level of cooperation just high enough to maintain the symbiont
cooperation. The differential rates of adaptation appear to lead
directly to the differential pay-offs inferred from studies of
host-endosymbiont systems.

\section{\label{sec:misers}
  Misers and Slaves}

\begin{figure}
  \includegraphics{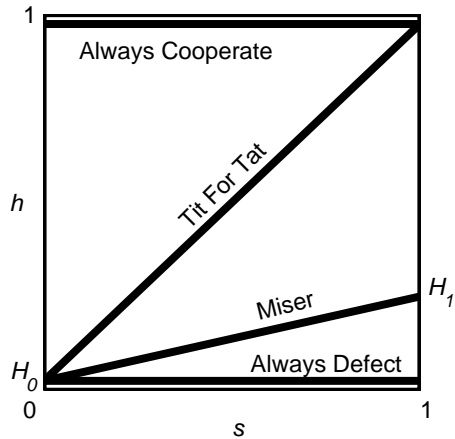}
\caption{\label{fig:one} Linear reactive strategies for \H. Each line is a possible mapping from $s$ to $h$, parameterized by $H_0$ and $H_1$.}
\end{figure}

In this section we consider just a pair of agents that interact, to
illustrate the idea that a slower rate of adaptation can be
advantageous. This appears closely related to the Red King effect
\cite{Bergstrom03}, although that model specifically avoids treating
the Prisoner's Dilemma scenario, in which there is a constant
temptation to defect on cooperative agreements. In later sections we
apply this notion to a model of endosymbiosis.

Consider two agents \S\ and \H, later we will interpret these as
symbiont and host.  Suppose that \S\ adopts some behaviour to a degree
$s$ at a cost to itself of $c_s$ per unit, and a benefit to \H\ of
$b_{s\rightarrow h}$ per unit. For example, this could be nitrogen
fixation by \emph{Rhizobia} which makes nitrogen available to legumes
in a useful form, but which entails a metabolic cost to the
\emph{Rhizobia}. Suppose that \H\ can also adopt some behaviour $h$
between zero and one. This benefits \S\ by $b_{h\rightarrow s}$ at a
cost of $c_h$ per unit. The net payoff to \S\ from the association is
then
\begin{equation}
  \label{eq:ws}
  w_s = b_{h\rightarrow s} \, h \; - \; c_s \, s
\end{equation}
while that to \H\ is
\begin{equation}
  \label{eq:wh}
  w_h = b_{s\rightarrow h} \, s \;  - \; c_h \, h
\end{equation}
For the purposes of our model we take $s$ and $h$ to be bounded below
by zero and above by one. That is, we assume there is some natural
limit to the amount of help that one organism can provide another.  If
benefits exceed costs for both parties, the maximum total payoff,
$w_s+w_h$, is achieved when both $s$ and $h$ are at their maximum
values. However any non-zero $s$ or $h$ only benefits the other party,
and so is an altruistic act of cooperation. By reducing the level of
cooperation, an agent reduces the costs, and so there is an incentive
for each agent to defect. This selfish behavior results in the payoff
to each agent decreasing and we have what is known in game theory
terms as a Prisoner's Dilemma \cite{Axelrod84}.

To provide an incentive for \S\ to cooperate, \H's behaviour needs to
depend on $s$ in some way, and the simplest non-trivial contingency is
linear, $ h = H_0 + (H_1 - H_0)s$. Here $H_0$ is the response to
complete defection ($s = 0$) and $H_1$ is the response to complete
cooperation ($s = 1$). Parameters $H_0$ and $H_1$ lie between zero and
one, ensuring that $h$ also lies in this range. The difference $H_1 -
H_0$ can be thought of as the `responsiveness' of \H. Completely
unresponsive agents have $H_0 = H_1$ , extreme cases being the naive
cooperator Always Cooperate (1,1) and the stalwart defector Always
Defect (0,0). For unresponsive agents, $h$ is independent of $s$ and
the gradient $\partial w_s /\partial s$ is $-c_s$ , so \S\ is
motivated to decrease its level of cooperation. In contrast, the
Tit-for-Tat strategy ($H_0$, $H_1$) = (0,1) is as responsive as
possible (see fig.~\ref{fig:one}). In this case we have $h=s$ and
$\partial w_s/\partial s = b_{h\rightarrow s} - c_s > 0$, so \S\ 
should cooperate more to increase its payoff. Notice that this
argument does not depend on specifying the details of \S's strategy
(meaning the way it arrives at the value $s$, given $h$) - it merely
indicates whether \S\ should increase or decrease the level of
cooperation it adopts with \H\ in pursuit of greater payoffs.

\begin{figure}
  \includegraphics{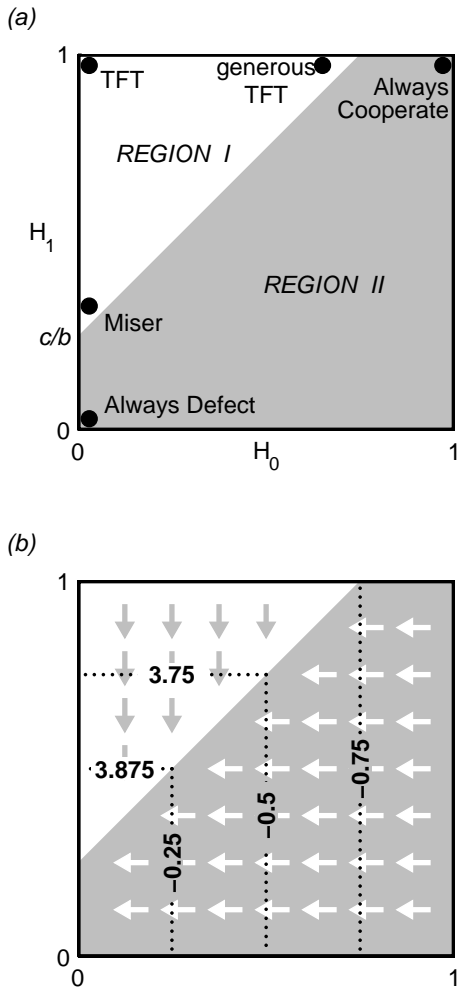}
\caption{\label{fig:two} (a)  The space of strategies for \H. Inside
  region I (unshaded) we have $H_1 - H_0 > c_s/b_{h\rightarrow s}$,
  which means that the optimal reply strategy is complete cooperation.
  In region II (shaded) the opposite holds, leading to pure defection.
  The dots mark the positions of named Prisoner's Dilemma strategies.
  (b) The payoff for \H, plotted on the same axes, assuming that \S\ 
  is adaptive enough to adopt the optimal reply behaviour. Contours
  (dotted lines) show \H's payoff and arrows show the direction of
  steepest ascent. The numbers shown are for the specific case of
  $c=1$, $b=4$ for both parties. The point on this surface with the
  highest payoff corresponds to the miser strategy of (0,
  $(c_s/b_{h\rightarrow s})^+$).}
\end{figure}

More generally, substituting the expression for $h$ into the payoff
for \S\, and differentiating with respect to $s$, shows that \S\ is
motivated to cooperate as much as it can, provided that the following
`incentive condition' is met,
\begin{equation}
  \label{eq:incentive}
  H_1 - H_0 > {c_s  \over b_{h\rightarrow s}}.
\end{equation}
This defines two regions on a plot of $H_1$ versus $H_0$ (see
fig.~\ref{fig:two}).  If \H's responsiveness is too low, \S\ would
benefit from lowering its level of cooperation, while if \H's
responsiveness is above the threshold $c_s/b_{h\rightarrow s}$, \S\ 
should become as cooperative as it can. Now consider a scenario in
which \S\ adapts its strategy over time in pursuit of higher payoffs.
If the incentive condition (eq.~\ref{eq:incentive}) is not met, $s$
will decrease to zero. In particular, any flat response ($H_0 = H_1$)
by \H\ will lead to pure defection by \S, and consequently cannot
account for the persistence of a mutualistic association between them.
Conversely, if the incentive condition is met, $s$ will rise to one.

Now consider the effect of this on \H. The payoff to an \H\ agent
whose responsiveness is above the threshold will be $b_{s\rightarrow
  h} - c_h H_1$ (since $s=1$), compared to $-c_h H_0$ for an \H\ below
the threshold (since then $s=0$). Since benefits outweigh costs
$b_{s\rightarrow h} - c_h H_1 > 0$ and so all \H\ strategies above the
threshold out-compete those below it. The best possible strategy for
$H$ will be the one with lowest $H_1$ that still obeys the incentive
condition. This strategy has $H_0 = 0$ and $H_1$ just above
$c_s/b_{h\rightarrow s}$ (it needs to be just above to ensure a
strictly positive gradient for $s$).  Equivalently, the optimal
strategy for \H\ is to cooperate with \S\ at a level
\begin{equation}
  \label{eq:hmiser}
  h_{\mathrm{miser}} = \biggl( {c_s \over b_{h\rightarrow s}} \, s \biggr)^+
\end{equation}
where the superscript ($+$) means `just above'.  The conclusion is
that the best strategy to adopt when faced with an optimally adaptive
co-player is that of a `miser'. The miser reacts with just enough
cooperation to encourage the other agent to cooperate more.  With
$H_1$ at $c_s/b_{h\rightarrow s}$ the miser collects a payoff of
$b_{s\rightarrow h} - c_s c_h /b_{h\rightarrow s} $, which is not much
less than the maximum payoff of $b_{s \rightarrow h}$ obtainable by
pure defection against the sucker strategy Always Cooperate. By
contrast, the highly adaptive agent is induced to cooperate at the
maximal level and for this receives a payoff only marginally above
zero. If \H\ is able to adapt but does so at a significantly lower
rate than \S, we might expect a miser-slave relationship to arise and
be stable. This phenomenon has been noted in simulations of the
prisoner's dilemma, although in that case the model involved was not
amenable to analysis \cite{Doebli98}.

It might be argued at this point that the preceding model is biased,
in that the \H\ strategies are responsive whereas \S\ just assumes a
fixed level of cooperation. The \S's are not given access to the same
repertoire of strategies as the \H\ strategies, and so perhaps their
enslavement is due merely to this built-in deficiency. To show that it
occurs in the same way when \emph{both} players may change their
strategies, introduce the parameters $S_0$ and $S_1$. The level of
cooperation is $s = S_0 + (S_1 - S_0)h$ and we have a simple linear
dynamical system in which $h$ depends on $s$, which depends on $h$
itself. The dynamics have a single attractor that is approached
rapidly, regardless of whether $h$ and $s$ are updated synchronously
or asynchronously \cite{Frean96}. This point is
\begin{eqnarray}
  h^* =& \alpha[H_0 +S_0(H_1 - H_0)], \\
  s^* =& \alpha[S_0 +H_0(S_1 - S_0)],
\end{eqnarray}
where $\alpha = [1 - (H_1 - H_0)(S_1 - S_0)]^{-1}$. Geometrically, the
point ($h^*$, $s^*$) is simply the intersection of the two strategies
plotted on the unit square. On fig.~\ref{fig:one}, an \S\ strategy
maps $h$ back to $s$ and must be a line from the top of the figure to
its base, crossing the \H\ strategy at one point. Successive updates
move the joint state towards this intersection point and convergence
to equilibrium is rapid (e.g. of the order of 10 to 20 responses). The
final levels determine the payoff that agents receive, namely
\begin{eqnarray}
  w_h^* &=& b_{h\rightarrow s} s^* - c_h h^*, \label{eq:whstar}\\
 w_s^* &=& b_{s\rightarrow h} h^* - c_s s^*, \label{eq:wsstar}
\end{eqnarray}
and similarly for \S.  If the player \S\ evolves its strategy
sufficiently rapidly, then it is straightforward to show that the
incentive condition is unchanged by this extension.

\section{\label{sec:hands}Hosts and Symbionts} 

\begin{figure}
  \includegraphics{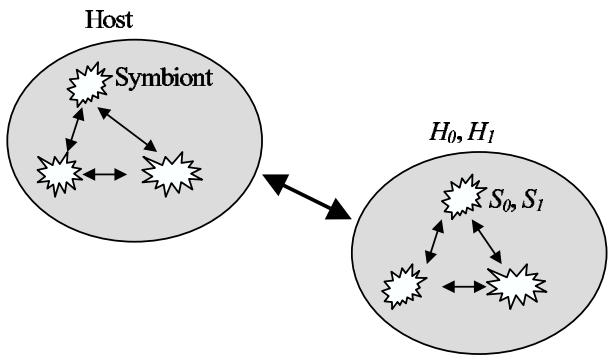}
\caption{\label{fig:three} 
  Hosts and their endosymbionts. Hosts compete with hosts, and
  symbionts compete with other symbionts in the same host, as
  indicated by the arrows.  }
\end{figure}
So far we have discussed just two agents and assumed that both adapt
their behaviour, with one of them (\S) adapting much more rapidly. It
is interesting to speculate whether this phenomenon might occur
between two co-evolving populations, and specifically between hosts
and their endosymbionts. In an endosymbiotic relationship, the
lifespan of the host is longer than that of the symbiont. Each of the
partners generates payoffs via interactions with the other type, yet
uses these to competitively replace only their own type. A single host
typically contains many endosymbionts, and these replicate within the
host for many generations. In the remainder of this paper we use
computer simulations to explore the idea that the preceding analysis
of agents \H\ and \S\ applies similarly to hosts and endosymbionts.
The intent is not to present a detailed model of a specific
host-endosymbiont system, but to examine in general terms whether this
mechanism can be applicable.

Our model (shown schematically in fig.~\ref{fig:three}) is intended to
capture the simplest aspects of endosymbiotic associations. We assume
purely horizontal transmission of symbionts, since this is the
situation that is most puzzling. Each host organism acquires
endosymbionts from the environment at some point in its early life,
and a symbiotic association is formed anew with each generation. The
host harbors its symbionts for some period, during which the symbionts
reproduce with mutations. We assume that symbionts compete for limited
resources within each host, leading to traits associated with greater
symbiont payoffs becoming more prevalent over time. Hosts are also in
a competitive environment, albeit one that occurs over a longer
time-scale. For simplicity we assume that variations in fitness are
due solely to the host-endosymbiont association.

\begin{figure}
  \includegraphics{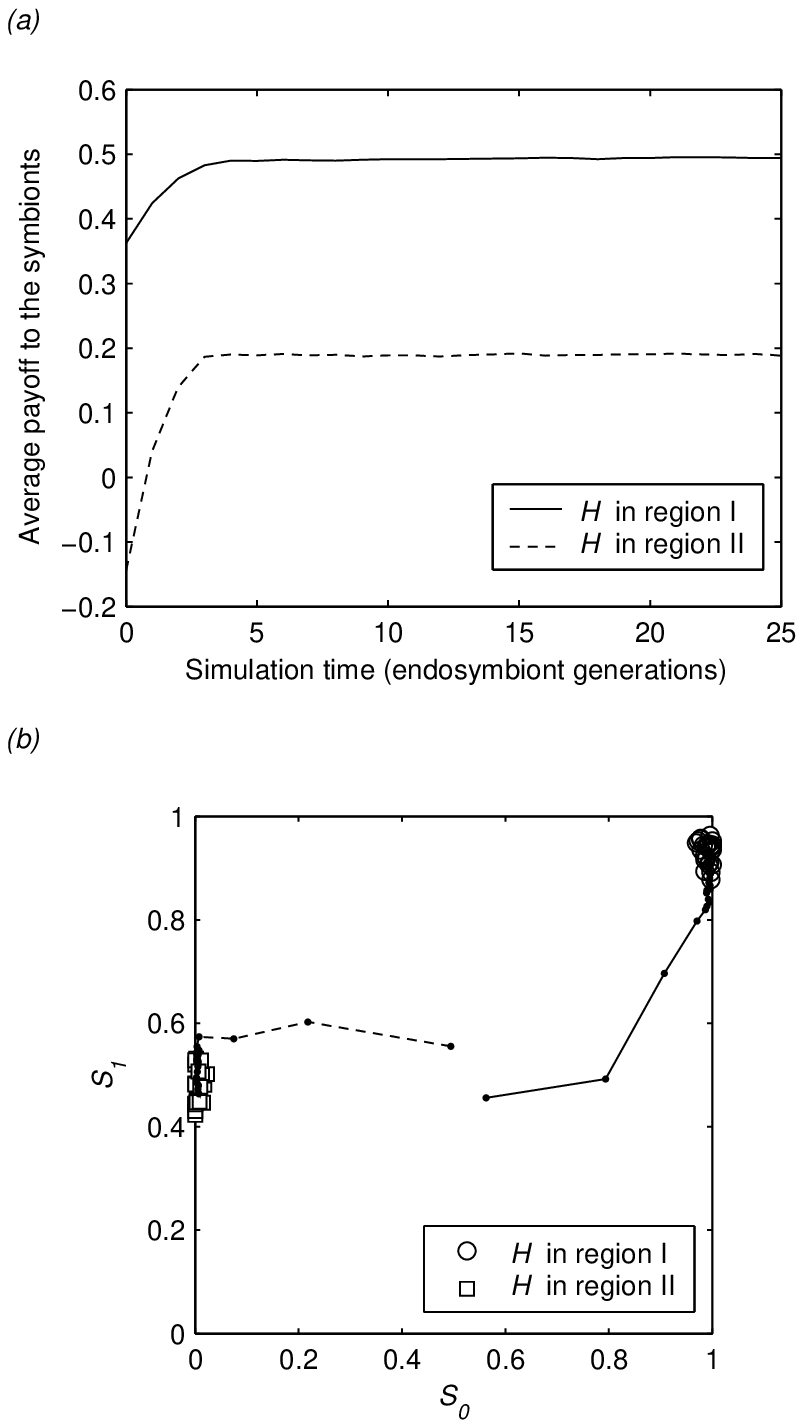}
\caption{\label{fig:four} 
  Examples of the evolution of endosymbionts within a single host. The
  solid lines and circles show results for symbionts within a host
  with a strategy in region I of fig.~(\ref{fig:two}), $H_0 = 0.02$,
  $H_1 = 0.15$. The dashed lines and squares show results for a
  symbionts associated with a host in region II, $H_0 = 0.02$, $H_1 =
  0.05$.  (a) A timeline of the average payoffs received by the
  symbionts. (b) The evolution of the symbiont strategies, showing the
  trajectory of the mean values of the parameters $S_0$ and $S_1$. The
  symbols show the strategies at the end of the 25 generation
  simulation. There is initially a rapid change in $S_0$, followed by
  a slower change in $S_1$.}
\end{figure}

Specifically, we model a population of 100 individuals of a host
species denoted \H, each with parameters ($H_0$, $H_1$). Each host has
an exclusive association with its own population of 25 symbionts,
denoted \S, and each such symbiont has a strategy ($S_0$, $S_1$). In
our simulations the population sizes are fixed: introducing
fluctuating populations would require additional assumptions (about
carrying capacities for instance).

All the evolvable parameters ($H_0$, $H_1$ for hosts and $S_0$, $S_1$
for symbionts) were started with random values between zero and one.
Each host applied its strategy of contingent cooperation in
interacting with each of its symbionts on an individual basis,
resulting in payoffs as described above. For simplicity, the
simulations shown here used the same costs and benefits for both types
of creature ($c_s = c_h$ and $b_{h\rightarrow s} = b_{s\rightarrow
  h}$). A very simple model of competitive replacement was used to
simulate the effect of natural selection, as follows. For symbionts,
two individuals were chosen at random from a single host. The payoff
to each individual was calculated using eqs.~(\ref{eq:wsstar}), but
with a small amount of random noise (uniformly distributed in the
range $\pm 0.01$) being added.
The symbiont with lower fitness (payoff) was then deleted and replaced by
the other. Rather than perfect replication of the better strategy,
mutations were introduced by adding random noise uniform in the range
$\pm 0.01$ to each parameter as it was copied, while constraining them
to remain within zero and one.  This allowed new strategies to arise
that were not present in the original population.  Exactly the same
procedure was applied to competition between hosts. We refer to $N$
replacements, where $N$ is the size of the relevant population, as one
generation. Endosymbionts underwent 25 such generations per host
replacement, and the simulation was run for 100 host generations. The
fitness of a given host was taken to be the sum of the payoffs from
its interactions with all its endosymbionts at the end of 100
endosymbiont generations. Each new host began life with randomly
generated new endosymbionts. This reflects the fact that the most
difficult case to understand is horizontal transmission, where symbionts exist in a free-living
form and infect hosts anew with each generation. Horizontal
transmission also appears to be the most common mode of maintaining
symbiont-host associations \cite{Maynardsmith95}.

\begin{figure}
  \includegraphics{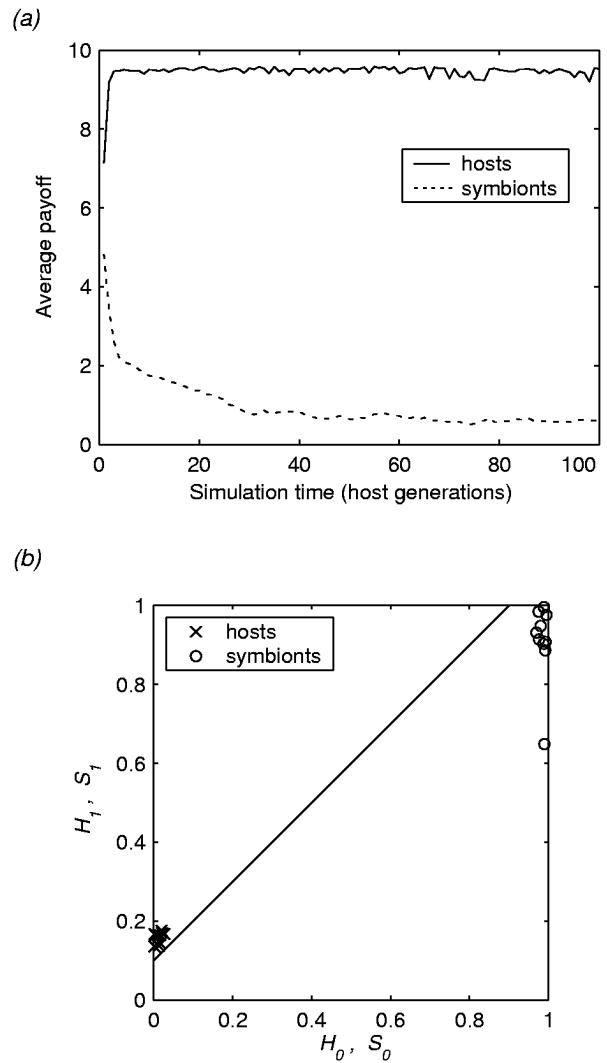}
\caption{\label{fig:five}
  Coevolution of hosts and endosymbionts.  (a) A timeline of the
  average payoffs received by hosts and symbionts (b) A random sample
  of the strategies present at the end of the simulation (to be
  compared with fig.~\ref{fig:two}).  }
\end{figure}
\begin{figure}
  \includegraphics{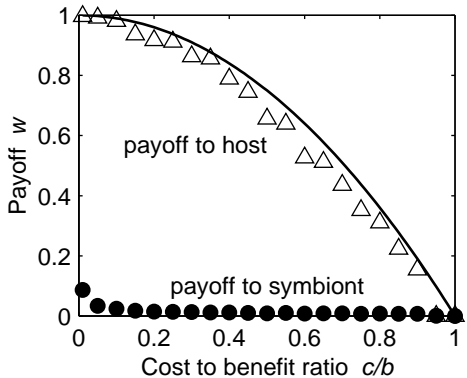}
\caption{\label{fig:six}The average payoffs to hosts (triangles) and
  symbionts (filled circles), versus the cost-benefit ratio, with $b =
  1$, where both species experience the same cost and benefits. The
  solid line indicates the predicted payoff for miserly hosts
  containing fully cooperative symbionts, $b - c^2/b$.  }
\end{figure}

Within each host, the symbionts evolve towards either full cooperation
or defection, depending on whether the host strategy is in region I or
II of fig.~\ref{fig:two}. The examples in fig.~\ref{fig:four} show the
evolution of the symbionts within two hosts, one slightly more
cooperative than the miser strategy, and one slightly less
cooperative. There is initially a rapid change in the symbiont
strategies, towards $S_0 = 1$ and $S_0 = 0$, respectively. The change
in $S_1$ is slower, because in the example given $H_0$ is small. As
expected, the payoff to the symbionts within a host increases with
time.

At the end of the full simulation the miser strategy dominated the
host population and most endosymbionts adopted strategies very close
to unconditional cooperation (see fig.~\ref{fig:five}). After an
initial transient, lasting for around 20 host generations, the average
payoff to the endosymbionts remained at less than one tenth of the
averaged payoff to the hosts. Note that the payoff to the symbionts
decreases as the hosts evolve. Occasionally a mutation would take a
host across the boundary into the region which is too non-cooperative,
this host's symbionts would then stop cooperating, the payoff to the
host would fall, and it would be eliminated from the population
through the selection process. The asymmetry in the rewards of the
association remains even at low cost-to-benefit ratios
(fig.~\ref{fig:six}). As $c$ approaches zero there is little cost to
cooperation and hence little incentive to defect, and yet the degree
of miserliness becomes increasingly severe.  Similar results are
achieved if the symbionts do not respond to the host (i.e. if the
constraint $S_0 = S_1$ is applied). In order to maintain the
cooperation of the symbionts the host strategy must be responsive. In
a simulation which imposed both $S_0 = S_1$ and $H_0 = H_1$ the hosts
and the symbionts both rapidly adopted strategies close to All Defect,
with little payoff to either party.  Similarly, if the host responds
not to the individual symbionts but rather to an averaged level of
cooperation, then mutual defection rapidly ensues.

\section{\label{sec:moderation}Three Moderating Influences} 

The argument given here makes a number of simplifying assumptions,
some of which accentuate the degree of miserliness. In real systems we
might expect somewhat less extreme outcomes for the following reasons.

Firstly, the model assumes that once symbionts enter a host they are
trapped there, at least until the host dies. Clearly things will be
different if they're free to go. One can draw the analogy with sets of
companies and their employees: for the worker there is a tradeoff
between staying (with poor but secure working conditions) and going in
search of greater prosperity with another employer, entailing some
risk associated with the transfer (i.e. the costs and risks of transit
itself, and the possibility that the new host/employer is even worse).
On the other hand, this implies some selection pressure in favour of
barriers or other `exit costs' that hosts might be expected to impose
on their workforces. For example, the `walling in' of \emph{Rhizobia}
in the root nodules of legumes may merely enhance the uptake of fixed
nitrogen by the plant, but it also could be argued that it plays a
role in preventing the bacteria from leaving.

Secondly, all hosts in the simulation contain the same number of
symbionts. One might instead expect the carrying capacity within more
miserly hosts to be lower, indirectly decreasing their fitness.  In
general, hosts should evolve to be as miserly as possible while still
retaining sufficient numbers of sufficiently compliant endosymbionts.

Finally, the model of selection employed here is a brutal one, and
this makes the division between regions I and II (fig.~\ref{fig:two})
particularly stark. As a host approaches the miser strategy the
gradient in fitness being followed by its endosymbiont population
becomes flatter and flatter, and at the miser strategy itself the
gradient is zero and so they spread out everywhere.  If the
endosymbionts aren't responding strongly to the host strategy, the
host's gradient disappears as well. Again we have a trade-off: a given
host wants to be as miserly as possible while ensuring that
differential incentives continue to encourage cooperation strongly
enough.

Against these points, however, the inheritance regime used in our
simulations makes it difficult for hosts to evolve cooperation in
their symbionts. Here each host begins life with an assortment of
endosymbionts with randomly chosen strategies. Increased levels of
symbiont cooperation are never passed on - they must be evolved anew
with each generation. More realistic models might allow the new
generation of hosts to begin with endosymbionts chosen from the
generation before. This could take the form of vertical inheritance
(in effect, perpetuating the `walling in' across generations), or else
symbionts might be expelled to the external environment and reacquired
by new hosts. In either case, it will be easier to evolve full
cooperation from symbionts, enhancing the tendency towards miserliness
in hosts.

\section{\label{sec:discussion}Discussion} 
A theme of recent work on symbiosis has been the surprising richness
and complexity of the entanglement between host and symbiont, in which
the partners undergo substantial metabolic and morphological changes
mediated by a complex series of mutual signals, even modulating one
another's gene expression \cite{Baumann98}.  It is clear that a model
of the simplicity described here ignores many potentially relevant
biological details. Its role is to show that, in principle, a low
pay-off by the host may be sufficient to coerce a population of
endosymbionts into nearly complete cooperation. An assumption of both
the theoretical analysis and the model is that the host is able to
respond to individual symbionts, preferentially rewarding those who
cooperate. Perhaps the complexity of the real-life symbiotic dialogue
concerns the host's efforts to detect and respond at a local scale to
the efforts of its symbiont passengers. The punishment of defecting
symbionts may not require the host to differentiate individual
symbionts, however. Suppose, for example, that a host species is well
served by symbionts that convert some compound $X$ into some other
compound $Y$. Even if the host is unable to reward the production of
$Y$ at a local scale, it can punish endosymbionts that don't carry out
the reaction by flooding the area with something that's toxic to them
in the presence of $X$, or toxic in the absence of $Y$. A mechanism
like this potentially allows a single host to interact with an
essentially arbitrary number of symbionts at the same time.

The theory proposed here accounts for both the persistence of
host-endosymbiont mutualisms and the observed asymmetry in their
benefits, as well as suggesting a role for the complex host structures
now being found in many associations. The very different dynamics
found in a two-species interaction, when compared with the usual
iterated Prisoner's Dilemma, suggests that there is still much to be
discovered about cooperative associations.

\begin{acknowledgments}
  The authors acknowledge the financial support of the Marsden Fund,
  which is administered by the Royal Society of New Zealand.
\end{acknowledgments}

\bibliography{fastslow}

\end{document}